\pdfoutput=1
\documentclass[10pt]{article}

\usepackage{jmlr2e}
\usepackage{amsmath}
\usepackage{microtype}
\usepackage{algorithm}
\usepackage{algorithmic}
\usepackage{multirow}
\usepackage{subfig}
\usepackage{bbold}
\usepackage{balance}
\usepackage{enumitem}
\usepackage{hyperref}

\newcommand{\Real}{\mathbb{R}}

\newcommand{\hide}[1]{}

\newcommand{\Ical}{\mathcal{I}}

\newcommand{\Scal}{\mathcal{S}}
\newcommand{\Ucal}{\mathcal{U}}
\newcommand{\Fcal}{\mathcal{F}}

\newcommand{\Rcal}{\mathcal{R}}
\newcommand{\Lcal}{\mathcal{L}}

\newcommand{\Tcal}{\mathcal{T}}

\newcommand{\V}{\mathbf{V}}
\renewcommand{\S}{\mathbf{S}}
\renewcommand{\r}{\mathbf{r}}
\newcommand{\vi}{\mathbf{v}}
\newcommand{\s}{\mathbf{s}}

\def\game{\textsc{FacT-CRS}}

\jmlrheading{1}{2000}{1-48}{4/00}{10/00}{A S M Ahsan-Ul Haque and Hongning Wang}

\ShortHeadings{Rethinking Conversational Recommendations: Is Decision Tree All You Need?}{Haque and Wang}
\firstpageno{1}

\begin{document}

\title{Rethinking Conversational Recommendations: Is Decision Tree All You Need?}

\author{\name A S M Ahsan-Ul Haque \email ah3wj@virginia.edu \\
	\addr Department of Computer Science\\
	University of Virginia\\
	Charlottesville, VA 22903, USA
	\AND
	\name Hongning Wang \email hw5x@virginia.edu\\
	\addr Department of Computer Science\\
	University of Virginia\\
	Charlottesville, VA 22903, USA
}

\maketitle

\begin{abstract}


Conversational recommender systems (CRS) dynamically obtain the users’ preferences via multi-turn questions and answers. The existing CRS solutions are widely dominated by deep reinforcement learning algorithms. However, deep reinforcement learning methods are often criticised for lacking interpretability and requiring a large amount of training data to perform.

In this paper, we explore a simpler alternative and propose a decision tree based solution to CRS.
The underlying challenge in CRS is that the same item can be described differently by different users. We show that decision trees are sufficient to characterize the \emph{interactions} between users and items, and solve the key challenges in multi-turn CRS: namely \textit{which questions to ask}, \textit{how to rank the candidate items}, \textit{when to recommend}, and \textit{how to handle user's negative feedback on the recommendations}. Firstly, the training of decision trees enables us to find questions which effectively narrow down the search space. Secondly, by learning embeddings for each item and tree nodes, the candidate items can be ranked based on their similarity to the conversation context encoded by the tree nodes. 
Thirdly, the diversity of items associated with each tree node allows us to develop an early stopping strategy to decide when to make recommendations. 
Fourthly, when the user rejects a recommendation, we adaptively choose the next decision tree to improve subsequent questions and recommendations. Extensive experiments on three publicly available benchmark CRS datasets show that our approach provides significant improvement to the state of the art CRS methods.

\end{abstract}



\section{Introduction}
\label{sec:intro}

Whereas traditional approaches in recommender systems infer user preferences solely based on historical user-item interaction data~\cite{he2017neural, wang2019neural, wang2019explainable}, conversational recommender systems (CRS) elicit user preferences through interactive question answering~\cite{salton1971smart, zhang2018towards, radlinski2017theoretical, chu1998evidential}. The traditional approaches are insufficient in adapting to the user's ongoing information need, since the user's preference can deviate over time from historical data. Moreover, users may look for different items under different situations. For example, a user who is looking for restaurant recommendations may seek a specific type of food at a particular time, or may consider other factors such as wait-time, availability of parking and etc differently when making her choices. A CRS solution (often referred to as ``agent'') can profile the user's current preference by the feedback collected from its strategically planned questions ~\cite{young2013pomdp, mccarthy2010experience}. 

Earlier forms of CRS can be traced back to interactive recommender systems~\cite{HE20169,Chen_Dai_Cai_Zhang_Wang_Tang_Zhang_Yu_2019, wu-etal-2019-proactive} and critiquing-based recommender systems~\cite{tversky1993context,tou1982rabbit,smyth2003analysis}. In the CRS setting proposed by Christakopoulou et al.~\cite{christakopoulou2016towards}, the agent searches for the target item using a multi-armed bandit algorithm solely in the item space, i.e., keep recommending different items. Zhang et al.~\cite{zhang2018towards} expanded the domain to the attribute space so that the agent needs to predict two things: what questions to ask and then which items to recommend, i.e., a series of questions followed by a recommendation. Li et al.~\cite{li2018towards} further expanded the setting by deciding between asking a question or making a recommendation at each round of interaction. As a result, the agent can ask a series of questions about item attributes and make several rounds of recommendations in an interleaved manner. This is often referred to as the multi-turn CRS setting and also the focus of our work.

\begin{figure}
    \centering
    \includegraphics[width = \linewidth]{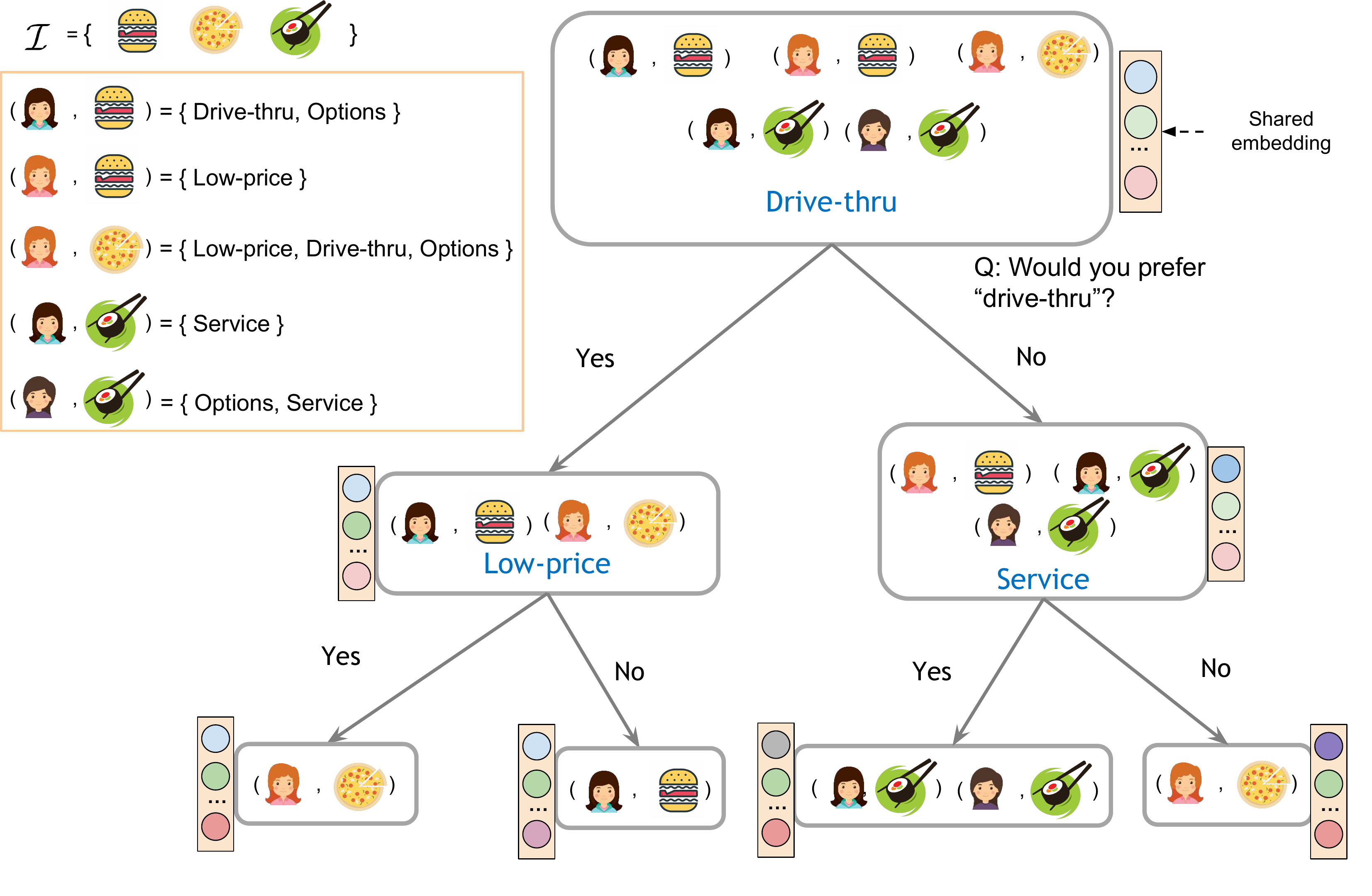}
    \caption{Overview of user-item interaction tree in \game{}. We can learn shared embedding of user-item interaction using the interaction tree and use it to ask questions and recommend. }
    \label{fig:example}
\end{figure}

Current multi-turn CRS solutions are dominated by deep reinforcement learning algorithms~\cite{earlei2020estimation, unicorndeng2021unified,scprlei2020interactive,fpanxu2021adapting}. Although encouraging empirical performance has been reported in  literature, we question whether a simpler alternative exists and can achieve comparable or even better performance, as the deep models are often criticized for their lack of interpretability and high demand of training data to perform. 
Intuitively, decision tree is a natural choice to construct the questions in multi-turn CRS \cite{tao2019fact,zhou2011functional}, where each question narrows down the candidate set of items. 
However, naively using a decision tree to partition the item space for the purpose of CRS is infeasible, because different users can view the same item differently and thus provide distinct answers to the same question even when they are looking for the same item. 
Figure~\ref{fig:example} shows this issue with an example in restaurant recommendation. From the observed user-item interactions, we notice that for the same burger shop, one user would describe it by its ``low price'' aspect, while another user did not pay attention to its price aspect but focuses on its ``options'' and ``drive-through'' service. 
As the partition of item space is exclusive in a decision tree, i.e., the items are separated into non-overlapping groups at the same level of tree nodes, if we used the question ``price'' to locate the restaurant in our interactions with the two users, we will surely fail to find the burger shop for at least one of them.

The problem described above motivates us to instead \emph{partition the user-item interactions}: the matching between a user-item pair during the course of CRS can be characterized by how the user would describe the item. As a result, a path on the decision tree groups the common descriptions about items from different users, and different descriptions of the same item then result in different paths on the tree. As shown in Figure \ref{fig:example}, the burger shop is now placed in different tree nodes to reflect the fact that different users would describe it differently.
This sheds light on the possibility of using a decision tree for multi-turn CRS.

In this paper, we propose a rule-based solution to multi-turn CRS, namely \game{} (Factorization Tree based Conversational Recommender System). We use user-item interactions to guide us to build a set of decision trees, which will be referred to as user-item interaction trees. Whereas the existing CRS methods~\cite{earlei2020estimation, unicorndeng2021unified, scprlei2020interactive, fpanxu2021adapting} rely on user embeddings to make recommendations, we propose to construct user-item interaction embeddings.
There are still three main challenges remaining to address in order to complete the solution. 
Firstly, \textit{how to rank the items?}  Clearly a decision tree naturally forms the questions. However, as shown in Figure~\ref{fig:example}, each node in the decision-tree may contain a varying number of items. 
When the number of candidate items is more than what we can recommend in a turn, we need to decide the ranking of items before making a recommendation. 
We extend the concept of factorization tree \cite{tao2019fact} to learn embeddings for user-item interactions and all the candidate items while constructing the decision trees. For all observed user-item interaction pairs located in a tree node (i.e., the users provided the same answers to the questions asked so far about their intended items), an embedding vector is assigned to match against the item embeddings. During training, the embedding vectors are learnt in a way such that the matched items will be ranked higher for the associated user-item pairs than those unmatched ones. At inference time, when we decide to make a recommendation at a tree node, we will use the corresponding interaction embedding to rank all candidate items. 

Secondly, \textit{when to make a recommendation?} 
It is desirable that the agent makes a recommendation as soon as it is confident, before the user’s patience wears out. It motives us to make a recommendation before exhausting the questions in the path of the interaction tree. An appropriate turn to make a recommendation is when 1) asking further questions does not provide much information gain, and 2) when the number of candidate items is small enough.
Based on those considerations, we propose two strategies to make early recommendations. When building the interaction tree, we keep track of how much information gain is achieved using the Gini Index and we stop splitting a node if the information gain is below a threshold. At inference time, we can make recommendations before reaching the leaf node, when the number of unique items associated on the node is less than a threshold.

Thirdly, \textit{how to handle a user's negative feedback about the recommended items?} 
Online feedback is an important aspect of multi-turn CRS where the agent should improve its action during a conversation, not only when the the user answers the planned question but also when he/she rejects a set of recommendations.  When the agent encounters a rejection, it becomes apparent that 1) the rejected items are not what the user is looking for, and 2) the target item is still on the lower part of the ranked list. Based on this insight, we design an adaptive feedback module to refine the inferred interaction embeddings based on the rejected items before moving to the next round of interaction.

To evaluate the effectiveness of \game{}, we conducted extensive experiments on three benchmark CRS datasets: LastFM, BookRec and MovieLens. The experiment results demonstrate that \game{} performed significantly better in finding the target items for the users using fewer turns. Extensive analysis shows that \game{} improved the conversation
by narrowing down the candidate items taking user-item interaction into account and adapting to the online feedback effectively.

\section{Related Works}
\label{sec:realted}

CRS leverages multiple turns of questions and answers to obtain user preferences in a dynamic manner.
There have been mainly four lines of research in CRS. In the item-based approaches, the agent keeps making recommendations based on users' immediate feedback. Christakopoulou et al. \cite{christakopoulou2016towards} proposed this line of research which marked the inception of CRS. They employed multi-armed bandit models to acquire the users’ feedback on individual items, 
such that model updates itself at each turn. 

Later in the question-driven setting, the domain was expanded so that the system needed to predict two things: 1) what questions about item attributes to ask, and 2) which items to recommend. Note that, in this setting the system only recommends by the end of conversation. 
Zhang et al.~\cite{zhang2018towards} proposed a model which consisted of three stages.  In the initiation stage, user initiates the conversation. In the conversation stage, the system asks about the user preferences on attributes of items. And in the final stage, the system recommends the items. Zou et al.~\cite{zou2020towards} proposed Qrec which asks questions in a predetermined number of times and then makes recommendation once. 
In Qrec, they chose the most uncertain attribute to ask (the attribute that the system has the least confidence between positive and negative feedback). Christakopoulou et al.~\cite{christakopoulou2018q} later studied the setting where the user can provide multiple answers (for example: ``choose all the attributes that you like."). There is another approach in CRS, which synthesize natural language. This adds personalized response to the conversation and it is applicable to an even broader scope. Usually, in this approach, there is a natural language understanding module and a natural language generation module.  This approach is beyond the scope of our work.

In this paper, we focus on the multi-turn setting of CRS, where an agent needs to choose between asking a question or making a recommendation in each turn of conversation. Lei et al.~\cite{earlei2020estimation} expanded the single-turn recommendation CRS to the multi-turn setting, where multiple questions and recommendations can be made in one conversation until the user accepts the recommendation or until the end of the conversation.

Reinforcement learning (RL) has been widely adopted in multi-turn CRS, which formulates the CRS problem as a Markov Decision Process (MDP)~\cite{pei2019value, shani2005mdp, zhao2018recommendations}. Recent works on RL-based interactive recommendation~\cite{wang2020kerl, xin2020self, zhang2019text, zhou2020interactive, zou2020pseudo} have been shown to effectively recommend items by modeling the users' dynamic preferences. The objective of these methods is to learn an effective RL policy to decide which items to recommend. Sun et al.~\cite{sun2018conversational} used a belief tracker based on an LSTM model to determine when to recommend, but their model was not able to handle when user rejected a recommendation.  Lei et al.~\cite{earlei2020estimation} utilized three different modules: the estimation module predicts user preference on  items and attributes; the action module decides whether to ask about attributes or to recommend items; and the reflection module updates the model when there is negative feedback. A dynamic preference model was introduced by Xu et al.~\cite{fpanxu2021adapting}, where they proposed a gating mechanism to include both positive and negative user feedback. Deng et al.~\cite{unicorndeng2021unified} combined the question selection and recommender modules. They proposed two heuristics to reduce the candidate action space by pre-selecting attributes and items in each turn. Zhang et al.~\cite{zhang2020conversational} used a bandit algorithm to select attributes and used a heuristic to decide whether to ask questions about attributes or make recommendations. Li et al.~\cite{li2021seamlessly} unified attributes and items in the same space in a multi-armed bandit setting to determine the questions, and used another bandit model to determine when to recommend. Li et al.~\cite{li2018towards} used a deep RL based model to decide when to make a recommendation or which question to ask. Chu et al. \cite{chu2022meta} developed a meta reinforcement learning based solution to handle new users in CRS. 

Although RL-based models dominate modern CRS solutions, in this paper, we explore a simpler and effective alternative based on decision trees. The RL methods provide strong baselines to compare the recommendation quality of our work.

\section{Methodology}
\label{sec:method}

In this section, we first describe the problem of multi-turn conversational recommendation. Our work is motivated by finding a simple decision tree based model to solve the challenges in multi-turn conversational recommendation. We first explain how a decision tree can effectively model the user-item interactions to capture the potential matching between a user and an item, which translates to a set of questions to filter and rank items for recommendations. Then we describe how to adapt the user-item interaction tree in \game{} to effectively address the challenges in CRS.

\subsection{Preliminary}

\textbf{Multi-turn CRS.} In this paper, we study the multi-turn conversational recommendation problem, since it has been popularly adopted because of its realistic setting~\cite{earlei2020estimation, unicorndeng2021unified, fpanxu2021adapting}. In this setting, the agent takes multiple turns to ask questions regarding item attributes or make recommendations (e.g., movies, restaurants etc.). We denote the set of items as $\Ical = \{i_1, i_2, \dots, i_n\}$, and the set of users as $\Ucal = \{u_1, u_2, \dots, u_m\}$. The set of attributes $\Fcal = \{f_1, f_2, \dots, f_p\}$ are used to describe the items. Each item is associated with a set of predefined attributes $\Fcal_i$.
Suppose that a user $u \in \Ucal$ is in a conversation with the CRS agent and her target item is $i^+ \in \Ical$. Each conversation constitutes of multiple turns. At each turn $t$, the agent needs to decide whether to ask a question or to make a recommendation. Depending on the agent's action, each turn can either be 1) a question asked by the agent and followed by the user’s answer, or 2) top-K recommendations followed by the user's acceptance or rejection. In the question-answer turn, the agent asks an attribute $f_t \in \Fcal$, i.e., ``do you prefer attribute $f_t$?'' In the recommendation turn, the user accepts the recommendation, if the target item $i^+$ is contained in the top-K recommendations. Otherwise, the user rejects the recommendation.  
The conversation is considered successful, if the user accepts a recommendation. Otherwise, the conversation fails, if the agent reaches a maximum turn limit. The goal of CRS is to successfully recommend items to the user with the fewest number of turns.

\textbf{User-item interaction.}  We use the pair $(u, i)$ to denote that user $u$ interacted with item $i$ in history. We use $\Rcal$ to denote the set of user-item interactions. Let the number of interaction $|\Rcal| = q$. Each user-item interaction $\r_{u, i}\in \{0,1\}^{p}$, where $(u,i) \in \Rcal$. Here, 1 denotes that the attribute is mentioned in the interaction between the user and item (e.g., the user uses this attribute to describe the item), and 0 denotes that the attribute is not mentioned.  An example could be: user $u$ describes item $i$ using the terms of $\{f_1, f_2\}$, as shown in Figure \ref{fig:example}. Here, the interaction content $\r$ would be a $p$ dimensional vector description with 1 for $f_1$, and $f_2$ and 0 for the remaining attributes.

\subsection{User-item Interaction Tree}
\label{sec:review-tree}

User-item interaction tree (or just ``interaction tree'' for short) is the building block of our \game{} model that we use for asking questions, narrowing down the candidate set of items, and ranking the candidate items. 
The goal of the interaction tree is to allow us to hierarchically learn the interaction embeddings as a function of the attributes $\Fcal$. We use FacT \cite{tao2019fact} to associate each interaction and each item with a $d$-dimension vector. The embeddings of all the interactions and items are respectively denoted as $\S \in \Real^{q \times p}$ and $\V \in \Real^{n \times p}$. Formally, we associate each interaction $(u, i) \in \Rcal$ with an embedding $\s_{u, i} \in \Real^d$ and each item $i$ with an embedding $\vi_i \in \Real^d$. Intuitively, we want that a) the dot product $\s_{u,i}^T\vi_i$ is maximized and b) $\s_{u,i}^T \vi_j$ is minimized when $j \neq i$. We achieve this using the cross-entropy loss defined in the following,
\begin{equation}
\label{eq:cross-entropy}
\begin{aligned}
\mathcal{L}_{C E}(\mathbf{S}, \mathbf{V}, \mathcal{R}) &=\sum_{(u, i) \in \mathcal{R}}\log\left(1-\sigma\left(\mathbf{s}_{u, i}^{T} \mathbf{v}_{i}\right)\right) 
\end{aligned}
\end{equation}

Here, $\sigma(.)$ is the sigmoid function. To differentiate the relative ordering among candidate items, we also introduce the Bayesian Pairwise Ranking (BPR) objective~\cite{rendle2012bpr}. The intuition is that the items that a given user has interacted with should be scored higher than the items the user did not interact with. We sample the negative set $\mathcal{D}_{u, i}^{n e g}$ for interaction $(u, i)$  by choosing $i_{neg}$ where $\left(u, i_{neg}\right) \notin \mathcal{R}$. The BPR loss is calculated as in Eq.~\eqref{eq:bpr},

\begin{equation}
\label{eq:bpr}
\mathcal{B}\left(\mathbf{s}_{u, i}, \mathbf{V}, \mathcal{D}_{u, i}^{n e g}\right)=\sum_{i_{\text {neg }} \in \mathcal{D}_{u, i}^{\text {neg }}} \log \sigma\left(\mathbf{s}_{u, i}^{T} \mathbf{v}_{i}-\mathbf{s}_{u, i}^{T} \mathbf{v}_{i_{\text {neg }}}\right)
\end{equation}

Each node $z$ in the interaction tree is associated with a question $f_z \in \Fcal$ (i.e., ``Do you prefer attribute $f_z$?''). Let us consider a subset of all interactions $\Rcal_z \subseteq \Rcal$. Based on the user-item interaction $(u, i)$ content $\r_{u,i}$ , we can check whether or not a specific attribute $f_l$  is mentioned. For each attribute $f_l$, we can partition the subset of user-item interactions  $\Rcal_z$ into two disjoint sets $\Rcal_{z_{+|f_l}}$ and $\Rcal_{z_{-|f_l}}$ as given in Eq.~\eqref{eq:partition},
\begin{equation}
\label{eq:partition}
\begin{aligned}
&\Rcal_{z_{+|f_l}}=\left\{(u, i)\mid r_{{u, i}_l} = 1 , \r_{u, i} \in \Rcal_{z}\right\} \\
&\Rcal_{z_{-|f_l}}=\left\{(u, i)\mid r_{{u, i}_l} = 0 , \r_{u, i} \in \Rcal_{z}\right\} \\
\end{aligned}
\end{equation}


All the user-item interactions in the same node share the same description path from the root node and thus will share the same interaction embedding. 
Let $\Rcal_{z_{+|f_l}}$ and $\Rcal_{z_{-|f_l}}$ be the observed user-item interactions in the resulting  partitions, and  $\s_{pos}, \s_{neg}$ be the corresponding user-item interaction embeddings in each of the partitions. Note that $\s_{pos}$ is the shared embedding of all the user-item interactions in $ \Rcal_{z_{+|f_l}}$ and $\s_{neg}$ is the shared embedding of all the user-item interactions in $\Rcal_{z_{-|f_l}}$. We can find both embeddings by solving the following optimization problem:

\begin{equation}
\label{eq:split}
\begin{aligned}
&\Lcal(f_l|\Rcal_z) = \min_{\s_{pos}, \s_{neg}, \V} \Lcal_{CE} (\s_{pos}, \V, \Rcal_{z_{+|f_l}}) + \Lcal_{CE} (\s_{neg}, \V, \Rcal_{z_{-|f_l}} ) \\
&+ \lambda_{BPR} \left(\sum_{(u,i) \in \Rcal_{z_{+|f_l}}}\mathcal{B}\left(\s_{u,i}, \V, D_{u,i}^{neg}\right)
 + \sum_{(u,i) \in \Rcal_{z_{-|f_l}}}\mathcal{B}\left(\s_{u,i}, \V, D_{u,i}^{neg}\right) \right)\\
 &+ \lambda_s \left(\|\s_{pos}\|_2 +  \|\s_{neg}\|_2 \right)
\end{aligned}
\end{equation}
where $\lambda_{BPR}$ is a coefficient that controls the trade-off between cross-entropy loss and BPR loss, $\|.\|$ is the L2 regularization to reduce model complexity, and $\lambda_s$ is the regularization coefficient.

An optimal attribute split would partition the  user-item interaction subset $\Rcal_z$, where the embeddings in each disjoint group minimize the loss as given by Eq.~\eqref{eq:split}. We can find the optimal attribute $f_z$ by exhaustively searching through the attribute set $\Fcal$ using Eq.~\eqref{eq:optimal}:
\begin{equation}
\label{eq:optimal}
f_z = \arg \min_{f_l \in \Fcal} \Lcal(f_l|\Rcal_z)
\end{equation}

We build the interaction tree by recursively splitting each partition until the maximum depth $H_{max}$ is reached. Once constructed, the user-item interaction tree allows us to infer the interaction embedding for a new interaction at inference time by following the tree structure.

\textbf{Candidate items.} Each node in the user-item interaction tree maintains a subset of observed interactions. Since each interaction is a user-item pair, the items in the interaction set of that node form the candidate items for recommendation. Formally, given the subset of interaction $\Rcal_z$ at node $z$, the candidate set of items is given by:
\begin{equation}
\label{eq:candidate}
\Ical_{z} = \{i | (u, i) \in \Rcal_z\}
\end{equation}

\textbf{Recommendation score.} Suppose, after traversing the user-item interaction tree, the inferred interaction embedding is  $\s{}_t$. The recommendation score of each candidate item at turn $t$ can be readily computed as:
\begin{equation}
\label{eq:score}
    w^{\Ical}_t(i) = \s{}_t^T  \vi_i
\end{equation}

\subsection{\game{}: User-item Interaction Tree for Conversational Recommendation}

Successfully solving the multi-turn CRS problem requires addressing the following four problems: namely, 1) \textit{which questions to ask}, 2) \textit{how to rank the candidate items}, 3) \textit{when to recommend}, and 4) \textit{how to handle user's negative feedback on the recommendations}. Based on our construction of interaction tree described in Section \ref{sec:review-tree}, the first two questions are partially addressed. And in this section, we mainly focus on the remaining two key questions, based on  the structure of a decision tree.  

\subsubsection{Which questions to ask and how to rank the candidate items?}
\label{sec:rq1}

User-item interaction tree is learnt to optimize the questions to be asked based on user feedback, i.e., following the paths on the tree to narrow down the search space.
However, given the maximum depth of a trained decision tree is fixed, we cannot ask more than $H_{max}$  questions using one tree. For our model to generalize, we need to ask an arbitrary number of questions. Random forest~\cite{breiman2001random} is an ensemble learning method that provides a feasible way to solve this challenge by creating multiple trees. We build a random forest of $N$ user-item interaction trees. We can build the interaction trees in parallel so that each interaction tree in the forest considers a maximum of $f_{max}$ attributes where $f_{max} \leq q$. These attributes are randomly sampled from the complete attribute set $\Fcal$. We use $\tau_j$ to denote the $j^{th}$ interaction tree and $\Tcal$ to denote the set of all interaction trees, therefore an interaction forest. We will discuss how to leverage these multiple trees to 
realize multi-turn CRS in Section \ref{sec:rq3}. 

As each tree node is associated with an interaction embedding vector, once the agent decides to make a recommendation, it will use Eq \eqref{eq:score} to compute the ranking scores of all candidate items and return the top-K items as the recommendation of this turn.

\subsubsection{When to recommend?}
\label{sec:rq2}
The interaction tree  gives us a natural way to decide when to recommend, i.e., when reaching the leaf node of the tree. 
But this restricts us to ask at most $H_{max}$ questions before we can start a recommendation turn. However, an important goal of conversational recommender systems is to minimize the number of interactions with the user. In this case, it is desirable to make a recommendation when the agent is ``confident'' about the item to be recommended. Hence, the question becomes: how to make an early recommendation? We use the following two strategies to recommend early.

\begin{itemize}[wide = 1pt]
\item During training (pruning): When building each  user-item interaction tree, we stop splitting a node when the items in that node is ``homogeneous'', i.e., most of the interactions are about only a few items. We use Gini Index \cite{gini1921measurement} for this purpose. 
Suppose, for an internal node $z$ in an interaction tree, the set of items from the user-item interaction $\Rcal_z$ is $\Ical_z$ as given by Eq.~\eqref{eq:candidate}. 
We calculate Gini Index of  node $z$ as:
\begin{equation}
\label{eq:gini}
G_z=1-\left(\frac{|\Ical_{z}|}{|\Rcal_{z}|}\right)^{2}
\end{equation}

If the Gini Index of a node is greater than a predetermined threshold $\gamma$, we stop further splitting that node. 

\item During testing (EarlyRec):
\label{subsec:stop-testing}
At inference time, a small number of items in a node is a strong indication that it is time to recommend. If we encounter a node in the user-item interaction tree that has no more than $\eta$ items, we can make an early recommendation. Let the set of items in that node be $\Ical_z$. Then, if $|\Ical_z| \leq \eta$, we make an early recommendation. 
If strictly $|\Ical_z| \leq K$, then we can include all the items in $\Ical_z$ in the recommendation. 
We rank all the items $\in \Ical \backslash \Ical_z$ using the scoring function in Eq.~\eqref{eq:score}. The remaining $K - |\Ical_z| $ items with the highest scores are included in the top-K recommendation.
\end{itemize}

\subsubsection{How to handle online negative feedback?}
\label{sec:rq3}
Handling negative feedback or recommendation rejection is an important challenge in CRS. When the user rejects a recommendation, it provides us a strong signal to improve the subsequent questions and recommendations. 
To effectively handle negative feedback from a user's  rejected recommendations, we approach this problem in three steps: 
retaining information from previously asked user-item interaction trees, effectively choosing the next user-item interaction tree, and updating the predicted user-item interaction embedding. 

\begin{itemize}[wide=0pt]
    \item {Retaining information from the previously asked user-item interaction trees.}
We need to move to a new user-item interaction tree when the user rejects our recommendation.

Suppose, at turn $t$, we have asked questions from the set of interaction trees $\Tcal^\alpha_t$, and we have obtained the set of user-item interaction embeddings $\Scal^{\alpha}_{t}$.
We keep the conversation history by updating the inferred interaction embedding to be the mean of the visited user-item interaction embeddings $\Scal^{\alpha}_{t}$. As a result, the updated interaction embedding at turn $t$ becomes:
\begin{equation}
\label{eq:mean-embedding}
\s_t = \frac{1}{|\Scal^{\alpha}_{t}|}\sum_{\s_j \in \Scal^{\alpha}_{t}} \s_j
\end{equation}

\item {Choosing the next user-item interaction tree.}
Assume that we have finished traversing an interaction tree and made a recommendation. If the user rejects, how do we select the next tree? One way is to randomly pick an unvisited tree. However, we believe that we can do better by using the following strategy. Since the predicted interaction embedding is expected be close to the target interaction embedding, we propose to move to the closest tree first. Suppose, at turn $t$, we have asked questions from  the set of $\Tcal^\alpha_t$ user-item interaction trees, and have asked the set of attributes $\Fcal^{\alpha}_{t} \subseteq \Fcal$. Based on Eq.~\eqref{eq:mean-embedding}, let the current inferred user-item interaction embedding be $\s_{t}$. We traverse the remaining  user-item interaction trees  in parallel by using only the attributes in $\Fcal^\alpha_{t}$, i.e, the attributes we already know the answers to. Assume that we get the interaction embedding $s'_j$ from the remaining interaction tree $\tau_j \in \Tcal \backslash \Tcal^{\alpha}_t$. We score each of the remaining tree $\tau_j$ according to how similar the corresponding interaction embedding $\s'_j$ is to the current interaction embedding $\s'_{t}$, as given by Eq.~\eqref{eq:tree-score},
\begin{equation}
\label{eq:tree-score}
w^{\Tcal}_t(\tau_j) = {\s'_j}^T \s{}
\end{equation}

Using this closest tree first strategy, we choose the user-item interaction tree $\tau_j$ with the highest similarity score as the next tree to continue the conversation.


\item{Updating the predicted user-item interaction embedding.}
We first identify the set of items that cause the failure and then update the inferred interaction embedding accordingly. Assume at turn $t$ we make a recommendation using the interaction embedding $\s_t$, which the user rejected.  We denote this set of recommended items as $\Ical_r$. 

We use the following two strategies to handle online feedback: 1) As $\Ical_r$ was rejected by the user, we update the predicted user-item interaction embedding to penalize the items in $\Ical_r$. 2) Since the target item is still expected to have high scores, we promote the items in the next top-K set. Based on $\s_t$, denote the next $K$ highest scoring items as set $\Ical_p$.

Combining both strategies, we update the user-item interaction embedding after a rejected recommendation as follows:
\begin{equation}
\label{eq:negative-feedback}
\s_{t+1} = \s_{t} +  \frac{\alpha_p}{|\Ical_p|} \sum_{i \in \Ical_p} \vi_i-  \frac{\alpha_n}{|\Ical_r|} \sum_{i \in \Ical_r} \vi_i
\end{equation}

Here, $\alpha_r$ and $\alpha_p$ are hyper-parameters that respectively determine how much we penalize the items in $\Ical_{r}$ and how much we promote the next top-ranked items $\Ical_p$. Algorithm~\ref{alg:forest} presents the inference steps of a single conversation session.

\end{itemize}

\begin{algorithm}
\caption{Inference in \game{}}\label{alg:forest}
  \begin{algorithmic}[1]

\REQUIRE Interaction forest $\Tcal$, interaction content $\r_{u, i} \in \{0,1\}^p$;
\ENSURE Outcome of conversation, either Success or Failure;
\STATE Current tree $\Tcal_t$ selected via cross-validation;
\STATE  Visited trees $\Tcal^\alpha_t = \{\}$;
\STATE Asked attributes $\Fcal^{\alpha}_{0} = \{\}$;
\STATE Inferred interaction embeddings $\Scal^{\alpha}_0 = \{\}$;
\FOR{turn $t = 1, 2, \dots, T$}
\STATE From $\Tcal_t$ ask $f_z$, and get $\Ical_z$, $\s_z$;
\WHILE{$f_z$ in $\Fcal^{\alpha}_t$}
\STATE From $\Tcal_t$ ask $f_z$, and get $\Ical_z$, $\s_z$;
\ENDWHILE
\STATE $\Fcal^{\alpha}_{t}  = \Fcal^{\alpha}_{t-1} \cup \{f_z\}$;
\IF{$|\Ical_z| \geq \eta$ and $z$ is not leaf node}
\STATE \textbf{continue};
\ENDIF
\STATE Get $\s_t$ from $\Scal_{t} = \Scal^{\alpha}_{t-1} \cup \{\s_z\}$ via Eq.~\eqref{eq:mean-embedding};
\STATE $\Ical_r =$ top-K items using $\s_t$ and $\Ical_z$ via Eq.~\eqref{eq:score};
\IF{User accepts $\Ical_r$}
\RETURN Success;
\ELSE
\STATE Update $\s_{t}$ via Eq.~\eqref{eq:negative-feedback};
\FOR {each remaining tree $\tau_j \in \Tcal\backslash \Tcal^\alpha_t$}
\STATE answer questions in $\Fcal^{\alpha}_t$ and get $\s'_j$;
\ENDFOR
\STATE $\Tcal_{t+1} = $ highest scoring tree via Eq.~\eqref{eq:tree-score};
\ENDIF
\ENDFOR
\RETURN Failure
  \end{algorithmic}
\end{algorithm}

\section{Experiments}
\label{sec:experiments}

To evaluate the effectiveness of \game{} in solving the challenges in CRS, we perform quantitative experiments guided by the following research questions (RQ):

\begin{itemize}[leftmargin = 10 pt]
    \item \textbf{RQ1.} Can a rule-based method i.e., \game{}, achieve better performance than the state-of-the-art RL-based methods in the multi-turn CRS setting?
    \item \textbf{RQ2.} How much improvement does early recommendation strategy offer compared to asking questions from the entire interaction tree?
    \item \textbf{RQ3.} Is online update of inferred interaction embedding useful in making better recommendations?
\end{itemize}

\subsection{Dataset}
\label{subsec:dataset}

We evaluate the recommendation quality of \game{} on three benchmark datasets used in multi-turn CRS. 

\begin{itemize}[leftmargin = 10pt]
    \item \textbf{LastFM}~\cite{bertin2011million} is a dataset for music artist recommendation. Lei et al.~\cite{earlei2020estimation} pruned the users with fewer than 10 reviews~\cite{he2017neural,rendle2012bpr} to reduce data sparsity, and  processed the original attributes by combining synonyms and removing low frequency attributes.  They categorized the original attributes into 33 coarse-grained attributes. 
\item \textbf{BookRec}~\cite{nguyen2019building} This is a book recommendation dataset filtered by removing low frequency TF-IDF attributes and keeping the top 35 attributes.

\item \textbf{MovieLens}~\cite{harper2015movielens} is a movie recommendation dataset filtered by keeping top 35 attributes according to their TF-IDF scores.
\end{itemize}

The statistics of the datasets are  shown in Table~\ref{table:dataset}. We randomly split the users into disjoint groups of 8:1:1 for training, validation and testing users. The code and data used in the experiments are available at \href{https://github.com/HCDM/XRec}{https://github.com/HCDM/XRec}.



\begin{table}[H]
\caption{Summary of datasets.}
\label{table:dataset}
\centering
\begin{tabular}{lrrr}
\hline & LastFM   & BookRec & MovieLens \\ 
\hline \# Users & 1,801  & 1,891 & 3,000 \\ 
\# Items & 7,432   & 4,343 & 5,974 \\ 
\# Interactions & 72,040 & 75,640 & 120,000 \\  
\# Attributes & 33 & 35 & 35 \\ 
\hline
\end{tabular}
\end{table}

\subsection{Experiment Settings}
\subsubsection{User simulator}
Conversational recommendation is a dynamic process for user preference elicitation. Similar to \cite{sun2018conversational, zhang2018towards, unicorndeng2021unified, earlei2020estimation, scprlei2020interactive, fpanxu2021adapting}, we created a user-simulator to enable the CRS training and testing. 
In each conversational session, an observed user-item pair $(u,i)$ is first selected. We call the item $i$ the target item or the ground-truth item for that conversation. Previous simulators~\cite{unicorndeng2021unified, earlei2020estimation, scprlei2020interactive, fpanxu2021adapting} assume that all of item $i$'s attributes $\mathcal{F}_i$ is the oracle set of attributes preferred by the user in this session. This means that all users will respond in the same way to the selected item. This setting is, however, unrealistic, because in reality, every user may not equally value every attribute of an item. Hence, this design eliminates the potential of personalized responses.

We design a user-based simulator that can handle user-specific feedback in each conversation round in the following way. Each item $i$ is associated with a set of attributes $\Fcal_i$, and each user $u$ has a preferred attribute set $\Fcal_u$. $\Fcal_i$ and $\Fcal_u$ are uniformly sampled for each item $i$ and user $u$ before the simulation starts. For a user-item interaction pair $(u,i)$, our simulator only accepts (responds ``Yes" to) an attribute $f_l$ if and only if it is mentioned in $\Fcal_i \cap \Fcal_u$; otherwise it will respond ``No''.
We set the maximum turn limit in a conversation to 10. The user leaves the conversation after the  turn limit is reached. We set K = 10, so that we are limited to recommend only 10 items in a recommendation turn.

\subsubsection{Evaluation Metrics}
\label{sec:metric}
Following~\cite{sun2018conversational, zhang2018towards, unicorndeng2021unified, earlei2020estimation, fpanxu2021adapting, scprlei2020interactive}, we use success rate and average turn as evaluation metrics. We use the success rate at turn T (SR@T) to measure the ratio of successful conversations. In an interaction where user $u$ interacted with item $i^+$, we call $i^+$ the ground-truth or the target item. A session of conversation is successful if the agent can identify the ground-truth item. We also report the average turns (AT) needed to end the round of conversation. The number of turns in a failed conversation is set to the the maximum turn limit T. The quality of recommendation is greater for larger SR@T, whereas the conversation is more efficient and to the point for smaller AT.

\subsubsection{Implementation Details.} We performed the training of \game{} on the training users, and tuned the hyper-parameters of our model on the validation set of users. The best model was chosen based on the validation success rate. The testing users were used to obtain the final reported performance for comparison. We fixed the embedding dimension $d = 40$, $\lambda_{BPR} = 10^{-3}$, $\alpha_p = 10^{-3}, \alpha_n = 10^-2$, and $G_z = 0.996$. The depth of the interaction tree was chosen as 7.

\subsubsection{Baselines}
\label{sec:baselines}
We evaluated the performance of \game{} and compared with the following state-of-the-art multi-turn CRS baselines~\cite{wu2015probabilisticmaxentropy, unicorndeng2021unified, earlei2020estimation, fpanxu2021adapting, scprlei2020interactive}.

\begin{itemize}[leftmargin = 10 pt]
\item \textbf{Max Entropy (MaxE)} \cite{wu2015probabilisticmaxentropy}: In this method, the CRS agent chooses either an attribute to ask or top ranked items to recommend in a probabilistic way. The agent asks the attribute with the maximum entropy based on the past conversation history. 
    
\item \textbf{EAR} \cite{earlei2020estimation}: This is a three stage approach for multi-turn CRS: the estimation stage builds a predictive model to estimate user preference based on both items and attributes, the action stage learns a policy to decide whether to ask about attributes or make a recommendation, and reflection stage updates the recommendation model based on online user feedback. 
    
\item \textbf{FPAN}~\cite{fpanxu2021adapting}: This extends EAR by dynamically revising user embeddings based on users’ feedback. The relationship between attribute-level and item-level feedback signals are used to identify the specific items and attributes that causes the rejection of an item. 
\begin{table*}[t]
    \centering
    \caption{Performance Comparison of different CRS models on three datasets. * represents the best performance among the baselines. The improvement over baseline is calculated against the best baseline values.}
    \label{tab:comparison}
    \begin{tabular}{lcccccc| c | c}
    \hline & & MaxE & EAR & FPAN & SCPR & UNI & \game{} &  Improvement*\\
    \hline \multirow{2}{*}{ LastFM } & SR@10 & $0.137$ & $0.428$ & $0.508^{*}$ & $0.432$ & $0.441$ & $0.719$ & 41.53\%\\
    & AT & $9.71$ & $8.62$ & $8.08^{*}$ & $8.70$ & $8.52$ & $6.65$ & 17.69\%\\
    \hline \multirow{2}{*}{ BookRec } & SR@10 & $0.206$ & $0.320$ & $0.397^{*}$ & $0.329$ & $0.358$ & $0.438$ & 10.33\%\\
    & AT & $9.64$ & $9.01$ & $8.31^{*}$ & $9.11$ & $9.00$ & $8.23$ & 0.96\%\\
    \hline \multirow{2}{*}{ MovieLens } & SR@10 & $0.262$ & $0.552$ & $0.589$ & $0.545$ & $0.596^{*}$ & $0.692$ & 16.10\%\\
    & AT & $9.46$ & $7.98$ & $7.81^{*}$ & $7.89$ & $8.01$ & $6.57$ & 15.87\%\\
    \hline
    \end{tabular}
\end{table*}

\item \textbf{SCPR}~\cite{scprlei2020interactive}:  It models CRS as an interactive path reasoning problem
on a knowledge graph. It leverages target user's preferred attributes by following user feedback to traverse attribute graph. Using the knowledge graph enables it to reduce the search space of candidate attributes.

\item \textbf{UNICORN} \cite{unicorndeng2021unified}: This method integrates the conversation and recommendation components into a unified RL solution. UNICORN develops a dynamic weighted graph based RL method to learn a policy for selecting the action at each conversation turn. It pre-selects attributes and items to simplify the RL training. 
    
\end{itemize}

All these methods rely on reinforcement learning models and pretrained user embeddings. To adopt them to new users at testing time, we use  use the mean embedding of the train users.

\subsection{Overall Performance}

To answer RQ1, we first evaluate the recommendation quality of \game{} in terms of success rate (SR@T) and average turn AT. A good CRS agent should be able to realize items which are more relevant to a user’s preference and rank them higher in a result list.

We report the results of our experiments in Table~\ref{tab:comparison}. \game{} consistently outperformed all baselines by a good margin on all datasets. FPAN performed better than the others among the baselines. We note that although it uses the same general structure as EAR, FPAN is able to generalize better on the new user because it updates the user embeddings dynamically with the user provided positive and negative feedback on attributes and items by two gated modules, which enables adaptive item recommendation. 
Although EAR is effective in handling large action space in CRS, its performance is limited by the separation of its conversation and recommendation components. UNICORN performs relatively better in this case through its usage of action selection strategy.

However, since all the baselines rely on user-level embedding, they cannot perform well on new test users. By hierarchically clustering the user-item interactions and exploiting the related user-item interactions, the interaction embeddings estimated by \game{} more effectively capture the current preference of the user, even in new users. This enables its good empirical performance in our evaluation. 

\begin{figure}[ht]
    \centering
    \includegraphics[width = 0.49\linewidth]{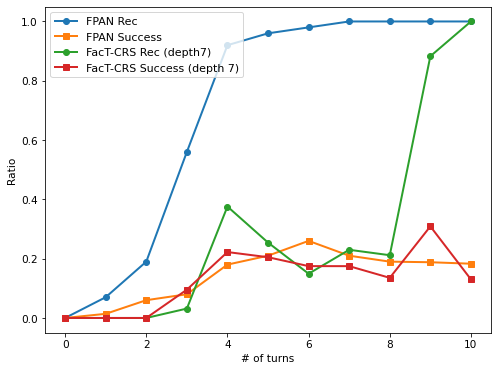}
    \includegraphics[width = 0.49\linewidth]{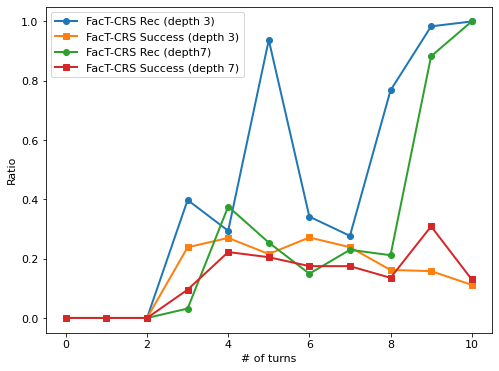}
    \caption{(Left) Ratio of recommendation and recommendation success rate at each turn on LastFM dataset (\game{} vs FPAN). (Right) Ratio of recommendation and recommendation success rate at each turn on LastFM dataset for \game{} at depth 3 and 7.}
    \label{fig:rec-succ}
\end{figure}

We also investigated how \game{} compares to other baselines at each turn of the conversations.  
To investigate this, we compared the recommendation probability and success rate between \game{} and FPAN on the LastFM dataset. As shown in Figure~\ref{fig:rec-succ}, FPAN can start recommending very early on, but when the user rejects a recommendation, FPAN tries to compensate by recommending more items, instead of identifying the cause of failure by asking more questions.

In comparison, \game{} handles the negative feedback  more effectively. It asks more clarifying questions before making another recommendation.
\game{} first tries to identify the negative items and then uses Eq.~\eqref{eq:negative-feedback} to update the predicted user-item interaction embedding, and then moves on to the next tree. This allows \game{} to subsequently ask better questions, and make better recommendations.

\subsubsection{Impact from tree depth.} Figure~\ref{fig:rec-succ} (right) shows the  performance of \game{} with different depths of its interaction trees. For smaller depth (at depth 3), \game{} starts recommending early and more frequently. However, at the same time it compromises the success rate. For larger depth (depth 7), \game{} asks more questions before making a recommendation. From the figure we see that the recommendation success rate improves by asking more questions.

\subsection{Ablation Study}
\label{sec:ablation}
In this section, we evaluate the effect of each individual component of \game{} by removing that component and evaluating the remaining model.

\subsubsection{Impact of Candidate Items in User-item Interaction Tree}

Every node in the user-item interaction tree is associated with a set  of user-item pairs. If we are ready to recommend at a node in the user-item interaction tree (usually near the leaf node), we check which items are in those user-item pairs. Then we rank those items based on the predicted user-item interaction embedding. This enables us to significantly narrow down the candidate set of items. Without this component, we would have to rank all the items each time using Eq.~\eqref{eq:score} to make a recommendation. Table~\ref{table:user-item-review-tree} reports our model's performance without taking into account the candidate items.
The candidate item set selection design is a vital part of \game{}. As Table~\ref{table:user-item-review-tree} shows, this is an effective way to make our recommendations much better by narrowing down the candidate set of items. Also, the early recommendation component of \game{} relies on the size of the candidate items to decide when to recommend.

\begin{table}[t]
\centering
\caption{Effect of Candidate Items in User-item Interaction Tree.}
\label{table:user-item-review-tree}
\setlength{\tabcolsep}{4pt}
\begin{tabular}{c|cc|cc|cc}
\hline
 & \multicolumn{2}{c|}{LastFM} & \multicolumn{2}{c|}{BookRec} & \multicolumn{2}{c}{MovieLens}\\
 & SR@10         & AT         & SR@10        & AT  & SR@10        & AT  \\
\hline \game{} & 0.719 & 6.65 & 0.438 & 8.23  & 0.692 & 6.57\\
\hline w/o Candidate & 0.578 & 8.36 & 0.195 & 9.50 & 0.493 & 8.51\\
\hline        
\end{tabular}
\end{table}

\begin{figure}[ht]
    \centering
    \includegraphics[width=0.49\linewidth]{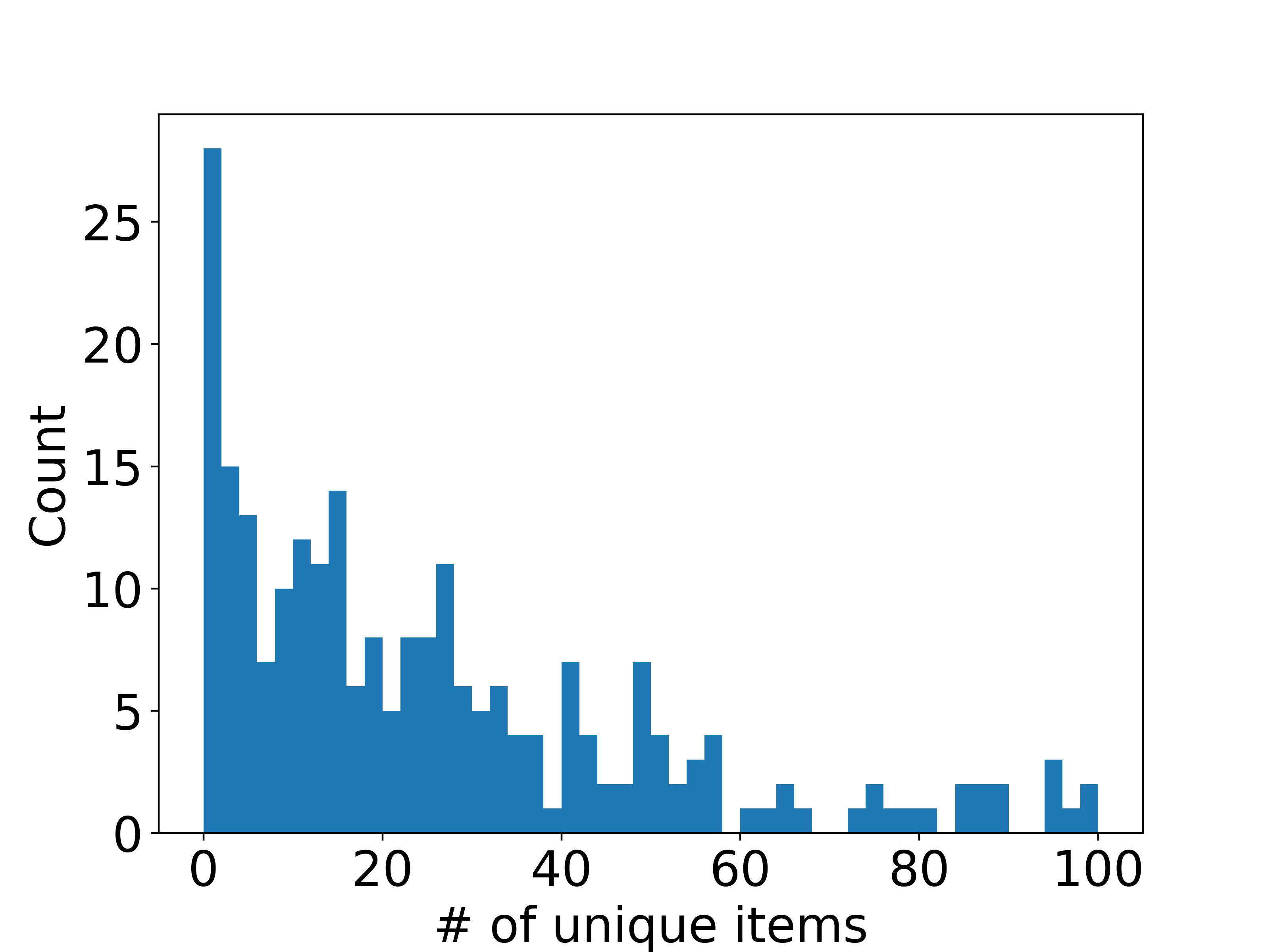}
    \includegraphics[width=0.49\linewidth]{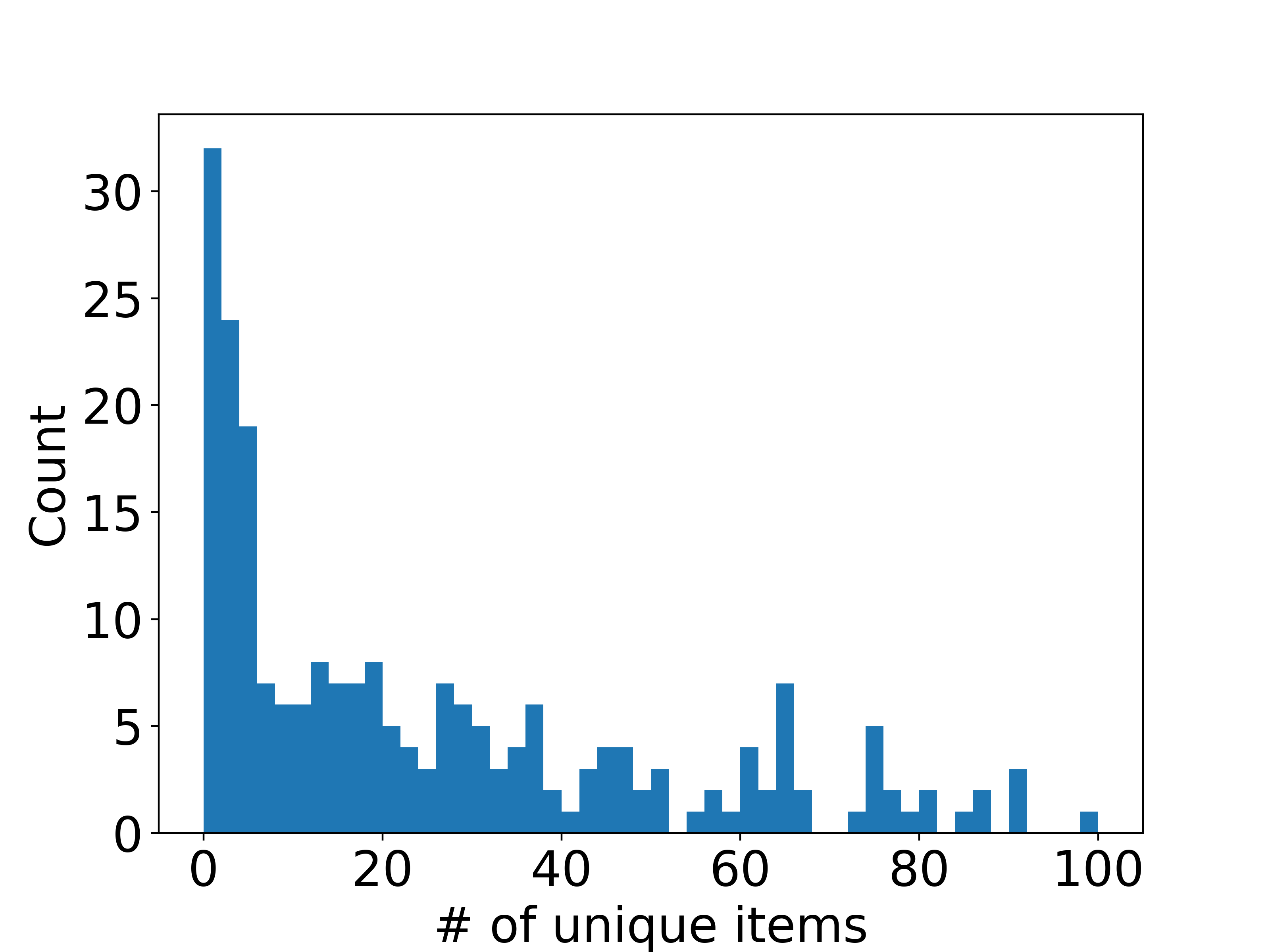}
    \caption{Histogram of the number of items in leaf nodes in a user-item interaction tree for (left) LastFM and (right) BookRec datasets.}
    \label{fig:hist-leaf-items}
\end{figure}

To understand why using the items in the leaf nodes of the user-item interaction tree is important, we plot the histogram of the number of unique items in the leaf node of a user-item interaction tree. Figure~\ref{fig:hist-leaf-items} shows the results for LastFM and BookRec datasets. As we can see, most of the leaf nodes have very few ($<20$) items. Hence, the leaf nodes work  well to cluster and narrow down the correct candidate list of items. This also empirically verifies our initial assumption that using the shared attributes to cluster the user-item interactions and subsequently learning the embeddings enhance the quality of the embeddings.

\begin{figure}[htb]
    \centering
    \includegraphics[width=0.49\linewidth]{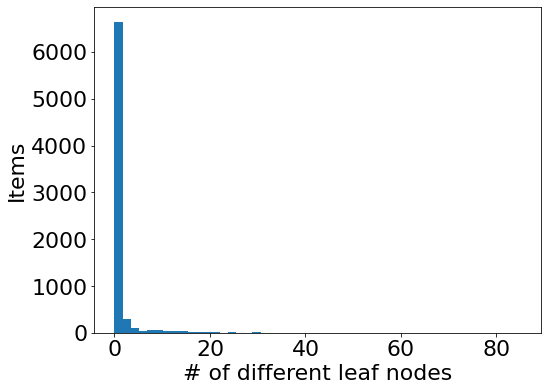}
    \includegraphics[width=0.49\linewidth]{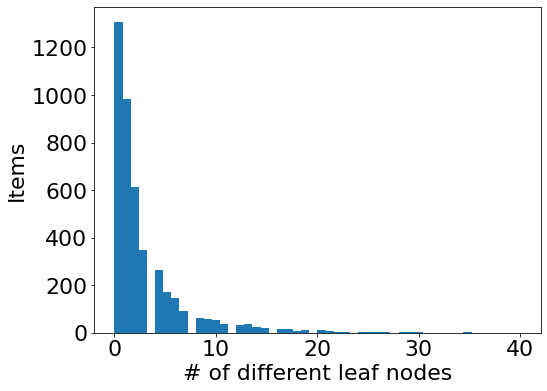}
    \caption{Histogram of the number of different leaf nodes each item appears in the user-item interaction tree for (left) Last FM and (right) BookRec datasets.}
    \label{fig:leaf_count_item}
\end{figure}

On the other hand, we also check how scattered each item is among the user-item interaction tree by recording how many different leaf nodes contain the same item. Figure~\ref{fig:leaf_count_item} shows the histogram of the number of leaf nodes each item is spread across. This figure demonstrates that the locations of most items are quite concentrated. By combining the findings from Figure~\ref{fig:hist-leaf-items} and \ref{fig:leaf_count_item}, we find that using the shared attributes to group items according to user-item interactions, \game{} can effectively find a good subset of candidate items containing a small number of items.

\begin{table}[t]
\centering
\caption{Ablation study of main components of \game{}.}
\label{table:ablation}
\setlength{\tabcolsep}{4pt}
\begin{tabular}{c|cc|cc|cc}
\hline
 & \multicolumn{2}{c|}{LastFM} & \multicolumn{2}{c|}{BookRec} & \multicolumn{2}{c}{MovieLens}\\
 & SR@10         & AT         & SR@10        & AT  & SR@10        & AT   \\
\hline \game{} &0.719 & 6.65  & 0.438 & 8.23 & 0.692 & 6.57\\
\hline  $\neg$ RF & 0.349 & 8.30 &  0.117 & 9.52 & 0.223 & 8.73\\
\hline  $\neg$ EarlyRec &  0.410 & 9.68  &  0.201 & 9.81 & 0.436 & 9.68\\
\hline $\neg$ OnlineFeed & 0.704 & 6.58 & 0.350 & 8.46 &  0.595 &  6.53 \\
\hline        
\end{tabular}
\end{table}

\subsubsection{Impact of Random Forest (RF)}
Random forest provides a key feature of multi-turn CRS by allowing us to ask more questions after encountering a rejection.
Without RF, the number of questions we can ask at most is the maximum depth of the tree $H_{max}$. We evaluate the significance of RF, by evaluating a single interaction tree built from the complete attribute set $\Fcal$.  Table~\ref{table:ablation} shows that it contributes the most among all components. By allowing \game{} to ask an arbitrary number of questions and to recommend multiple times, this component makes the user-item interaction tree suitable for multi-turn CRS setting.

\subsubsection{Impact of Early Recommendation (EarlyRec)}

For RQ2, we study the contribution of the EarlyRec strategy. To minimize the number of interactions with the user, we make an early recommendation if the candidate set of items at the current node is small enough. As we can see from Table~\ref{table:ablation}, EarlyRec has significant impact on minimizing the average turn (AT). By recommending early, this component also helps the agent understand if it is on the right track, and allows the agent to make necessary corrections that improve future recommendations.

\subsubsection{Impact of Handling Online Negative Feedback (OnlineFeed)}
To answer RQ3, we study the effectiveness of our online feedback method. The online feedback component updates the current inferred interaction embedding, when the user rejects recommendations made by the agent. As we can see from Table~\ref{table:ablation}, this strategy contributes to better recommendation quality. OnlineFeed component   first tries to identify the items responsible for the rejected recommendations, and corrects the predicted user-item interaction embedding to move towards the potential set of items that contains the target item. This allows \game{} to choose the next interaction tree more efficiently.

\subsection{Case Study}
\label{sec:case-study}

We performed the following case studies to analyze the performance of our model and to identify where we can further improve.

\subsubsection{Failed Conversations}
\label{subsec:failed-conversations}

We paid special attention to the failed conversations to get a better understanding of why a conversation fails. On all three datasets, we report the average number of mentioned attributes in the failed interaction and compare it to successful interactions. Table~\ref{table:failed-conversation} summarizes the mean and standard deviation of this results.
As we can see, the average number of mentioned attributes in the failed conversations is smaller than the average number of mentioned attributes in the corresponding complete datasets. 

\begin{table}[t]
\centering
\caption{Mean $\mu$ and standard deviation $\sigma$ of number of mentioned attributes in the interaction for successful, failed, and all conversations.}
\label{table:failed-conversation}
\begin{tabular}{c|cc|cc|cc}
\hline
\multicolumn{1}{c|}{} & \multicolumn{2}{c|}{LastFM} & \multicolumn{2}{c|}{BookRec} & \multicolumn{2}{c}{MovieLens}\\ 
                     & $\mu$       & $\sigma$    & $\mu$       & $\sigma$    & $\mu$       & $\sigma$  \\ \hline
Successful           & 5.65 & 1.13
 & 5.60 & 1.09 & 4.37 & 1.01\\ \hline
Failed               & 5.15 & 1.16 & 5.06 & 0.99 & 4.02 & 0.91\\ \hline
All                  & 5.51 & 1.15 & 5.30 & 1.07 & 4.26 & 1.00
\\ \hline
\end{tabular}
\end{table}

We next look at the conversations \game{} failed where the interaction contained at least $p_n$ number of  attributes in   Table~\ref{tab:min-attributes}. Our experiments show that for greater values of $p_n$, it becomes increasingly likely that \game{} can successfully recommend the target item. This gives us the basic insight of why a conversation fails. Since the attributes mentioned in the failed conversations are very few, our model can not find sufficient information to infer which particular item the user is looking for.

\begin{table}
\caption{SR@10 of \game{} when at least $p_n$ number of attributes confirmed in the interaction.}
\label{tab:min-attributes}
\centering
\begin{tabular}{c|c|c|c}
\hline
Min \# attributes $p_n$ & LastFM             & BookRec            & MovieLens          \\ \hline
3                 & 0.721 & 0.438 & 0.729  \\ \hline
4                 & 0.751  & 0.496 & 0.765 \\ \hline
5                 & 0.783   & 0.573 & 0.830  \\ \hline
6                 & 0.826 & 0.644 & 0.882 \\ \hline
\end{tabular}
\end{table}

\subsubsection{Identified Attributes}

\begin{figure}[t]
    \centering
    \includegraphics[width=0.49\linewidth]{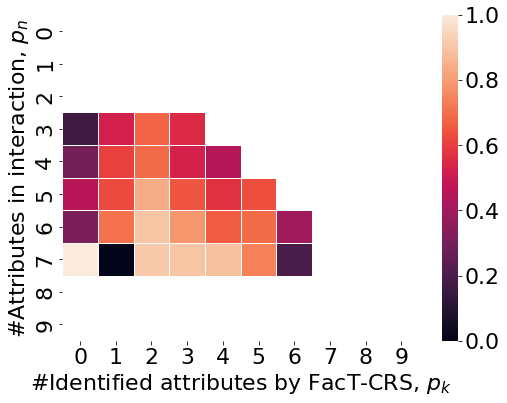}
    \includegraphics[width=0.49\linewidth]{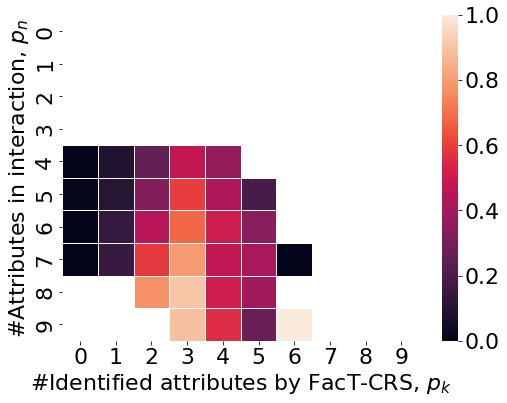}
    \caption{SR@10 of the number of attributes identified (correctly asked) by \game{} for different interaction length on (left) LastFM and (right) BookRec datasets.}
    \label{fig:heatmap}
\end{figure}

Figure~\ref{fig:heatmap} shows the success rate of conversations with different interaction length $p_n$ and the number of attributes $p_k$ identified by \game{}. Note that $ p_k > p_n$ is not possible, i.e., \game{} cannot identify more attributes than the total number of attributes associated with an interaction. Also, $p_k < T$, since any CRS agent can ask at most $T-1$ questions. When $p_k \leq p_n$, the white cells in Figure~\ref{fig:heatmap} refer to the events that are possible but did not occur. For example, on the BookRec dataset, when the number of attributes in an interaction is 9 (i.e., $p_n = 9$), \game{} always identified at least 3 attributes ($p_k \geq 3$). When more attributes are identified (left to right in Figure \ref{fig:heatmap}), it is more likely that the conversation will be successful. Similarly, when the user mentions more attributes in an interaction (top to bottom), \game{} is likely to identify more attributes and subsequently the conversations are more likely to be successful.

Our interaction tree based solution provides good predictive accuracy as well as minimizes the inputs required from the user.  Our model outperforms the existing RL-based baselines on LastFM, BookRec and MovieLens datasets, which demonstrates the robustness of our model.

\section{Conclusion}
\label{sec:conclusion}
Multi-turn CRS is a dynamic approach to elicit the current user preference by asking a series of questions and making recommendations accordingly. Existing approaches in conversational recommender system rely heavily on reinforcement learning based policy learning, whose performance however strongly depends on the amount of training data. 

In this paper, we proposed an alternative to the reinforcement learning methods and demonstrated multi-turn CRS are addressable by decision trees. To generalize a decision tree for multi-turn CRS, we addressed four key challenges in multi-turn CRS: \textit{which questions to ask, how to rank the items, when to recommend, and how to handle the user's rejection}. We proposed building an user-item interaction tree that is able to identify different description of a certain item. The interaction tree naturally provides a way to ask questions. To effectively rank the items, we learned the embeddings of user-item interactions. For this purpose, we used a decision tree based method called the factorization tree~\cite{tao2019fact}, which allows us to narrow down the candidate set of items by asking questions. By leveraging the random forest, we extended factorization tree to multi-turn CRS. We solved the challenge of when to recommend, by recommending when the candidate item set is small enough. Making corrections to the interaction embedding after encountering a rejection enables us to effectively handle the users' online rejection.
We extensively experimented on three benchmark CRS datasets, and compared \game{}'s performance with existing RL-based state-of-the-art solutions. The experimental results demonstrate that \game{} outperforms on all three datasets by successfully asking questions and identifying the target items in fewer number of turns.

Our exploration in \game{} sheds light on simple alternatives for multi-turn CRS. Though effective in our extensive evaluations, our solution still contains several empirically set hyper-parameters, such as the tree depth and number of trees. It is important for us to eliminate such hyper-parameters via automated tuning. In addition, currently we handle the conversations independently, even if they were from the same user. As our future work, it is important for us to study how to leverage observations from the same user or about the same item to further facilitate the conversation and recommendation. 
\section{Acknowledgement}

We thank the anonymous reviewers for their valuable comments. This work is partially supported by NSF under grant IIS-2128019, IIS-2007492 and IIS-1553568.

\balance
\bibliography{0.0.0ref}

\begin{thebibliography}{44}
\providecommand{\natexlab}[1]{#1}
\providecommand{\url}[1]{\texttt{#1}}
\expandafter\ifx\csname urlstyle\endcsname\relax
  \providecommand{\doi}[1]{doi: #1}\else
  \providecommand{\doi}{doi: \begingroup \urlstyle{rm}\Url}\fi

\bibitem[Bertin-Mahieux et~al.(2011)Bertin-Mahieux, Ellis, Whitman, and
  Lamere]{bertin2011million}
Thierry Bertin-Mahieux, Daniel~PW Ellis, Brian Whitman, and Paul Lamere.
\newblock The million song dataset.
\newblock 2011.

\bibitem[Breiman(2001)]{breiman2001random}
Leo Breiman.
\newblock Random forests.
\newblock \emph{Machine learning}, 45\penalty0 (1):\penalty0 5--32, 2001.

\bibitem[Chen et~al.(2019)Chen, Dai, Cai, Zhang, Wang, Tang, Zhang, and
  Yu]{Chen_Dai_Cai_Zhang_Wang_Tang_Zhang_Yu_2019}
Haokun Chen, Xinyi Dai, Han Cai, Weinan Zhang, Xuejian Wang, Ruiming Tang,
  Yuzhou Zhang, and Yong Yu.
\newblock Large-scale interactive recommendation with tree-structured policy
  gradient.
\newblock \emph{Proceedings of the AAAI Conference on Artificial Intelligence},
  33\penalty0 (01):\penalty0 3312--3320, Jul. 2019.
\newblock \doi{10.1609/aaai.v33i01.33013312}.
\newblock URL \url{https://ojs.aaai.org/index.php/AAAI/article/view/4204}.

\bibitem[Christakopoulou et~al.(2016)Christakopoulou, Radlinski, and
  Hofmann]{christakopoulou2016towards}
Konstantina Christakopoulou, Filip Radlinski, and Katja Hofmann.
\newblock Towards conversational recommender systems.
\newblock In \emph{Proceedings of the 22nd ACM SIGKDD international conference
  on knowledge discovery and data mining}, pages 815--824, 2016.

\bibitem[Christakopoulou et~al.(2018)Christakopoulou, Beutel, Li, Jain, and
  Chi]{christakopoulou2018q}
Konstantina Christakopoulou, Alex Beutel, Rui Li, Sagar Jain, and Ed~H Chi.
\newblock Q\&r: A two-stage approach toward interactive recommendation.
\newblock In \emph{Proceedings of the 24th ACM SIGKDD International Conference
  on Knowledge Discovery \& Data Mining}, pages 139--148, 2018.

\bibitem[Chu et~al.(2022)Chu, Wang, Xiao, Long, and Wu]{chu2022meta}
Zhendong Chu, Hongning Wang, Yun Xiao, Bo~Long, and Lingfei Wu.
\newblock Meta policy learning for cold-start conversational recommendation.
\newblock \emph{arXiv preprint arXiv:2205.11788}, 2022.

\bibitem[Chu-Carroll and Brown(1998)]{chu1998evidential}
Jennifer Chu-Carroll and Michael~K Brown.
\newblock An evidential model for tracking initiative in collaborative dialogue
  interactions.
\newblock In \emph{Computational Models of Mixed-Initiative Interaction}, pages
  49--87. Springer, 1998.

\bibitem[Deng et~al.(2021)Deng, Li, Sun, Ding, and Lam]{unicorndeng2021unified}
Yang Deng, Yaliang Li, Fei Sun, Bolin Ding, and Wai Lam.
\newblock Unified conversational recommendation policy learning via graph-based
  reinforcement learning.
\newblock \emph{arXiv preprint arXiv:2105.09710}, 2021.

\bibitem[Gini(1921)]{gini1921measurement}
Corrado Gini.
\newblock Measurement of inequality of incomes.
\newblock \emph{The economic journal}, 31\penalty0 (121):\penalty0 124--126,
  1921.

\bibitem[Harper and Konstan(2015)]{harper2015movielens}
F~Maxwell Harper and Joseph~A Konstan.
\newblock The movielens datasets: History and context.
\newblock \emph{Acm transactions on interactive intelligent systems (tiis)},
  5\penalty0 (4):\penalty0 1--19, 2015.

\bibitem[He et~al.(2016)He, Parra, and Verbert]{HE20169}
Chen He, Denis Parra, and Katrien Verbert.
\newblock Interactive recommender systems: A survey of the state of the art and
  future research challenges and opportunities.
\newblock \emph{Expert Systems with Applications}, 56:\penalty0 9--27, 2016.
\newblock ISSN 0957-4174.
\newblock \doi{https://doi.org/10.1016/j.eswa.2016.02.013}.
\newblock URL
  \url{https://www.sciencedirect.com/science/article/pii/S0957417416300367}.

\bibitem[He and Chua(2017)]{he2017neural}
Xiangnan He and Tat-Seng Chua.
\newblock Neural factorization machines for sparse predictive analytics.
\newblock In \emph{Proceedings of the 40th International ACM SIGIR conference
  on Research and Development in Information Retrieval}, pages 355--364, 2017.

\bibitem[Lei et~al.(2020{\natexlab{a}})Lei, He, Miao, Wu, Hong, Kan, and
  Chua]{earlei2020estimation}
Wenqiang Lei, Xiangnan He, Yisong Miao, Qingyun Wu, Richang Hong, Min-Yen Kan,
  and Tat-Seng Chua.
\newblock Estimation-action-reflection: Towards deep interaction between
  conversational and recommender systems.
\newblock In \emph{Proceedings of the 13th International Conference on Web
  Search and Data Mining}, pages 304--312, 2020{\natexlab{a}}.

\bibitem[Lei et~al.(2020{\natexlab{b}})Lei, Zhang, He, Miao, Wang, Chen, and
  Chua]{scprlei2020interactive}
Wenqiang Lei, Gangyi Zhang, Xiangnan He, Yisong Miao, Xiang Wang, Liang Chen,
  and Tat-Seng Chua.
\newblock Interactive path reasoning on graph for conversational
  recommendation.
\newblock In \emph{Proceedings of the 26th ACM SIGKDD International Conference
  on Knowledge Discovery \& Data Mining}, pages 2073--2083, 2020{\natexlab{b}}.

\bibitem[Li et~al.(2018)Li, Ebrahimi~Kahou, Schulz, Michalski, Charlin, and
  Pal]{li2018towards}
Raymond Li, Samira Ebrahimi~Kahou, Hannes Schulz, Vincent Michalski, Laurent
  Charlin, and Chris Pal.
\newblock Towards deep conversational recommendations.
\newblock \emph{Advances in neural information processing systems}, 31, 2018.

\bibitem[Li et~al.(2021)Li, Lei, Wu, He, Jiang, and Chua]{li2021seamlessly}
Shijun Li, Wenqiang Lei, Qingyun Wu, Xiangnan He, Peng Jiang, and Tat-Seng
  Chua.
\newblock Seamlessly unifying attributes and items: Conversational
  recommendation for cold-start users.
\newblock \emph{ACM Transactions on Information Systems (TOIS)}, 39\penalty0
  (4):\penalty0 1--29, 2021.

\bibitem[McCarthy et~al.(2010)McCarthy, Salem, and
  Smyth]{mccarthy2010experience}
Kevin McCarthy, Yasser Salem, and Barry Smyth.
\newblock Experience-based critiquing: Reusing critiquing experiences to
  improve conversational recommendation.
\newblock In \emph{International Conference on Case-Based Reasoning}, pages
  480--494. Springer, 2010.

\bibitem[Nguyen et~al.(2019)Nguyen, Rocco, and Ruscio]{nguyen2019building}
Phuong~T Nguyen, Juri~Di Rocco, and Davide~Di Ruscio.
\newblock Building information systems using collaborative-filtering
  recommendation techniques.
\newblock In \emph{International Conference on Advanced Information Systems
  Engineering}, pages 214--226. Springer, 2019.

\bibitem[Pei et~al.(2019)Pei, Yang, Cui, Lin, Sun, Jiang, Ou, and
  Zhang]{pei2019value}
Changhua Pei, Xinru Yang, Qing Cui, Xiao Lin, Fei Sun, Peng Jiang, Wenwu Ou,
  and Yongfeng Zhang.
\newblock Value-aware recommendation based on reinforcement profit
  maximization.
\newblock In \emph{The World Wide Web Conference}, pages 3123--3129, 2019.

\bibitem[Radlinski and Craswell(2017)]{radlinski2017theoretical}
Filip Radlinski and Nick Craswell.
\newblock A theoretical framework for conversational search.
\newblock In \emph{Proceedings of the 2017 conference on conference human
  information interaction and retrieval}, pages 117--126, 2017.

\bibitem[Rendle et~al.(2012)Rendle, Freudenthaler, Gantner, and
  Schmidt-Thieme]{rendle2012bpr}
Steffen Rendle, Christoph Freudenthaler, Zeno Gantner, and Lars Schmidt-Thieme.
\newblock Bpr: Bayesian personalized ranking from implicit feedback.
\newblock \emph{arXiv preprint arXiv:1205.2618}, 2012.

\bibitem[Salton(1971)]{salton1971smart}
Gerard Salton.
\newblock \emph{The SMART retrieval system—experiments in automatic document
  processing}.
\newblock Prentice-Hall, Inc., 1971.

\bibitem[Shani et~al.(2005)Shani, Heckerman, Brafman, and
  Boutilier]{shani2005mdp}
Guy Shani, David Heckerman, Ronen~I Brafman, and Craig Boutilier.
\newblock An mdp-based recommender system.
\newblock \emph{Journal of Machine Learning Research}, 6\penalty0 (9), 2005.

\bibitem[Smyth and McGinty(2003)]{smyth2003analysis}
Barry Smyth and Lorraine McGinty.
\newblock An analysis of feedback strategies in conversational recommenders.
\newblock In \emph{the Fourteenth Irish Artificial Intelligence and Cognitive
  Science Conference (AICS 2003)}. Citeseer, 2003.

\bibitem[Sun and Zhang(2018)]{sun2018conversational}
Yueming Sun and Yi~Zhang.
\newblock Conversational recommender system.
\newblock In \emph{The 41st international acm sigir conference on research \&
  development in information retrieval}, pages 235--244, 2018.

\bibitem[Tao et~al.(2019)Tao, Jia, Wang, and Wang]{tao2019fact}
Yiyi Tao, Yiling Jia, Nan Wang, and Hongning Wang.
\newblock {The FacT: Taming latent factor models for explainability with
  factorization trees}.
\newblock In \emph{Proceedings of the 42nd International ACM SIGIR Conference
  on Research and Development in Information Retrieval}, pages 295--304, 2019.

\bibitem[Tou et~al.(1982)Tou, Williams, Fikes, Henderson~Jr, and
  Malone]{tou1982rabbit}
Frederich~N Tou, Michael~D Williams, Richard Fikes, D~Austin Henderson~Jr, and
  Thomas~W Malone.
\newblock Rabbit: An intelligent database assistant.
\newblock In \emph{AAAI}, pages 314--318, 1982.

\bibitem[Tversky and Simonson(1993)]{tversky1993context}
Amos Tversky and Itamar Simonson.
\newblock Context-dependent preferences.
\newblock \emph{Management science}, 39\penalty0 (10):\penalty0 1179--1189,
  1993.

\bibitem[Wang et~al.(2020)Wang, Fan, Xia, Zhao, Niu, and Huang]{wang2020kerl}
Pengfei Wang, Yu~Fan, Long Xia, Wayne~Xin Zhao, ShaoZhang Niu, and Jimmy Huang.
\newblock Kerl: A knowledge-guided reinforcement learning model for sequential
  recommendation.
\newblock In \emph{Proceedings of the 43rd International ACM SIGIR conference
  on research and development in Information Retrieval}, pages 209--218, 2020.

\bibitem[Wang et~al.(2019{\natexlab{a}})Wang, He, Wang, Feng, and
  Chua]{wang2019neural}
Xiang Wang, Xiangnan He, Meng Wang, Fuli Feng, and Tat-Seng Chua.
\newblock Neural graph collaborative filtering.
\newblock In \emph{Proceedings of the 42nd international ACM SIGIR conference
  on Research and development in Information Retrieval}, pages 165--174,
  2019{\natexlab{a}}.

\bibitem[Wang et~al.(2019{\natexlab{b}})Wang, Wang, Xu, He, Cao, and
  Chua]{wang2019explainable}
Xiang Wang, Dingxian Wang, Canran Xu, Xiangnan He, Yixin Cao, and Tat-Seng
  Chua.
\newblock Explainable reasoning over knowledge graphs for recommendation.
\newblock In \emph{Proceedings of the AAAI conference on artificial
  intelligence}, volume~33, pages 5329--5336, 2019{\natexlab{b}}.

\bibitem[Wu et~al.(2015)Wu, Li, and Lee]{wu2015probabilisticmaxentropy}
Ji~Wu, Miao Li, and Chin-Hui Lee.
\newblock A probabilistic framework for representing dialog systems and
  entropy-based dialog management through dynamic stochastic state evolution.
\newblock \emph{IEEE/ACM Transactions on Audio, Speech, and Language
  Processing}, 23\penalty0 (11):\penalty0 2026--2035, 2015.

\bibitem[Wu et~al.(2019)Wu, Guo, Zhou, Wu, Zhang, Lian, and
  Wang]{wu-etal-2019-proactive}
Wenquan Wu, Zhen Guo, Xiangyang Zhou, Hua Wu, Xiyuan Zhang, Rongzhong Lian, and
  Haifeng Wang.
\newblock Proactive human-machine conversation with explicit conversation goal.
\newblock In \emph{Proceedings of the 57th Annual Meeting of the Association
  for Computational Linguistics}, pages 3794--3804, Florence, Italy, July 2019.
  Association for Computational Linguistics.
\newblock \doi{10.18653/v1/P19-1369}.
\newblock URL \url{https://aclanthology.org/P19-1369}.

\bibitem[Xin et~al.(2020)Xin, Karatzoglou, Arapakis, and Jose]{xin2020self}
Xin Xin, Alexandros Karatzoglou, Ioannis Arapakis, and Joemon~M Jose.
\newblock Self-supervised reinforcement learning for recommender systems.
\newblock In \emph{Proceedings of the 43rd International ACM SIGIR Conference
  on Research and Development in Information Retrieval}, pages 931--940, 2020.

\bibitem[Xu et~al.(2021)Xu, Yang, Xu, Gao, Guo, and Wen]{fpanxu2021adapting}
Kerui Xu, Jingxuan Yang, Jun Xu, Sheng Gao, Jun Guo, and Ji-Rong Wen.
\newblock Adapting user preference to online feedback in multi-round
  conversational recommendation.
\newblock In \emph{Proceedings of the 14th ACM international conference on web
  search and data mining}, pages 364--372, 2021.

\bibitem[Young et~al.(2013)Young, Ga{\v{s}}i{\'c}, Thomson, and
  Williams]{young2013pomdp}
Steve Young, Milica Ga{\v{s}}i{\'c}, Blaise Thomson, and Jason~D Williams.
\newblock Pomdp-based statistical spoken dialog systems: A review.
\newblock \emph{Proceedings of the IEEE}, 101\penalty0 (5):\penalty0
  1160--1179, 2013.

\bibitem[Zhang et~al.(2019)Zhang, Yu, Shen, Jin, and Chen]{zhang2019text}
Ruiyi Zhang, Tong Yu, Yilin Shen, Hongxia Jin, and Changyou Chen.
\newblock Text-based interactive recommendation via constraint-augmented
  reinforcement learning.
\newblock \emph{Advances in neural information processing systems}, 32, 2019.

\bibitem[Zhang et~al.(2020)Zhang, Xie, Li, and CS~Lui]{zhang2020conversational}
Xiaoying Zhang, Hong Xie, Hang Li, and John CS~Lui.
\newblock Conversational contextual bandit: Algorithm and application.
\newblock In \emph{Proceedings of The Web Conference 2020}, pages 662--672,
  2020.

\bibitem[Zhang et~al.(2018)Zhang, Chen, Ai, Yang, and Croft]{zhang2018towards}
Yongfeng Zhang, Xu~Chen, Qingyao Ai, Liu Yang, and W~Bruce Croft.
\newblock Towards conversational search and recommendation: System ask, user
  respond.
\newblock In \emph{Proceedings of the 27th acm international conference on
  information and knowledge management}, pages 177--186, 2018.

\bibitem[Zhao et~al.(2018)Zhao, Zhang, Ding, Xia, Tang, and
  Yin]{zhao2018recommendations}
Xiangyu Zhao, Liang Zhang, Zhuoye Ding, Long Xia, Jiliang Tang, and Dawei Yin.
\newblock Recommendations with negative feedback via pairwise deep
  reinforcement learning.
\newblock In \emph{Proceedings of the 24th ACM SIGKDD International Conference
  on Knowledge Discovery \& Data Mining}, pages 1040--1048, 2018.

\bibitem[Zhou et~al.(2011)Zhou, Yang, and Zha]{zhou2011functional}
Ke~Zhou, Shuang-Hong Yang, and Hongyuan Zha.
\newblock Functional matrix factorizations for cold-start recommendation.
\newblock In \emph{Proceedings of the 34th international ACM SIGIR conference
  on Research and development in Information Retrieval}, pages 315--324, 2011.

\bibitem[Zhou et~al.(2020)Zhou, Dai, Chen, Zhang, Ren, Tang, He, and
  Yu]{zhou2020interactive}
Sijin Zhou, Xinyi Dai, Haokun Chen, Weinan Zhang, Kan Ren, Ruiming Tang,
  Xiuqiang He, and Yong Yu.
\newblock Interactive recommender system via knowledge graph-enhanced
  reinforcement learning.
\newblock In \emph{Proceedings of the 43rd International ACM SIGIR Conference
  on Research and Development in Information Retrieval}, pages 179--188, 2020.

\bibitem[Zou et~al.(2020{\natexlab{a}})Zou, Chen, and Kanoulas]{zou2020towards}
Jie Zou, Yifan Chen, and Evangelos Kanoulas.
\newblock Towards question-based recommender systems.
\newblock In \emph{Proceedings of the 43rd International ACM SIGIR Conference
  on Research and Development in Information Retrieval}, pages 881--890,
  2020{\natexlab{a}}.

\bibitem[Zou et~al.(2020{\natexlab{b}})Zou, Xia, Du, Zhang, Bai, Liu, Nie, and
  Yin]{zou2020pseudo}
Lixin Zou, Long Xia, Pan Du, Zhuo Zhang, Ting Bai, Weidong Liu, Jian-Yun Nie,
  and Dawei Yin.
\newblock Pseudo dyna-q: A reinforcement learning framework for interactive
  recommendation.
\newblock In \emph{Proceedings of the 13th International Conference on Web
  Search and Data Mining}, pages 816--824, 2020{\natexlab{b}}.

\end{thebibliography}
\end{document}